\documentclass[letterpaper]{article} 
\usepackage[dvipsnames, svgnames, x11names]{xcolor}

\usepackage{graphicx}
\usepackage{float} 
\usepackage{subfigure}
\usepackage{amsmath}
\usepackage{multicol}
\usepackage{multirow}
\usepackage{booktabs}
\usepackage{bbding}
\usepackage{amssymb}
\usepackage{newfloat}
\usepackage{listings}
\usepackage{color}
\usepackage{amsthm}
\usepackage[algo2e]{algorithm2e} 
\newtheorem*{theorem*}{Theorem}

\usepackage{aaai25}  
\usepackage{times}  
\usepackage{helvet}  
\usepackage{courier}  
\usepackage[hyphens]{url}  
\usepackage{graphicx} 
\usepackage{natbib}  
\usepackage{caption} 
\usepackage{algorithm}
\usepackage{algorithmic}

\usepackage{newfloat}
\usepackage{listings}
\DeclareCaptionStyle{ruled}{labelfont=normalfont,labelsep=colon,strut=off} 
\floatstyle{ruled}
\newfloat{listing}{tb}{lst}{}
\floatname{listing}{Listing}
\title{Exploring Query Efficient Data Generation towards Data-free Model Stealing in Hard Label Setting}
\author{
    Gaozheng Pei\textsuperscript{\rm 1}, Shaojie lyu\textsuperscript{\rm 3}, Ke Ma\textsuperscript{\rm 1\footnote{Corresponding author}}, Pinci Yang\textsuperscript{\rm 1}, Qianqian Xu\textsuperscript{\rm 2$^*$}, Yingfei Sun\textsuperscript{\rm 1}\\
}
\affiliations{
    \textsuperscript{\rm 1}School of Electronic, Electrical and Communication Engineering, UCAS, Beijing, China\\

    \textsuperscript{\rm 2}
    Key Laboratory of Intelligent Information Processing, Institute of Computing Technology, CAS, Beijing, China\\
    \textsuperscript{\rm 3} Tencent Corporate

    peigaozheng23, yangpinci23@\{mails.ucas.ac.cn\} \quad shaojielv@tencent.com\\
    make, yfsun@\{ucas.ac.cn\} \quad xuqianqian@ict.ac.cn
}

\usepackage{bibentry}

\begin{document}

\maketitle

\begin{abstract}
Data-free model stealing involves replicating the functionality of a target model into a substitute model without accessing the target model's structure, parameters, or training data. The adversary can only access the target model's predictions for generated samples. Once the substitute model closely approximates the behavior of the target model, attackers can exploit its white-box characteristics for subsequent malicious activities, such as adversarial attacks. Existing methods within cooperative game frameworks often produce samples with high confidence for the prediction of the substitute model, which makes it difficult for the substitute model to replicate the behavior of the target model. This paper presents a new data-free model stealing approach called Query Efficient Data Generation (\textbf{QEDG}). We introduce two distinct loss functions to ensure the generation of sufficient samples that closely and uniformly align with the target model's decision boundary across multiple classes. Building on the limitation of current methods, which typically yield only one piece of supervised information per query, we propose the query-free sample augmentation that enables the acquisition of additional supervised information without increasing the number of queries.  Motivated by theoretical analysis, we adopt the consistency rate metric, which more accurately evaluates the similarity between the substitute and target models. We conducted extensive experiments to verify the effectiveness of our proposed method, which achieved better performance with fewer queries compared to the state-of-the-art methods on the real \textbf{MLaaS} scenario and five datasets.
\end{abstract}

%

\section{Introduction}
The widespread adoption of cloud services by Microsoft \cite{Azure}, Google \cite{Google}, and Amazon  \cite{amazon} suggest that machine learning as a service (\textbf{MLaaS}) \cite{DBLP:conf/icmla/RibeiroGC15} may become the predominant mode of interaction between human and models in the future. From the user's perspective, these models operate as black boxes, providing feedback based exclusively on user inputs. This interaction method safeguards the privacy of \textbf{MLaaS} training data and ensures the security of the model. This prompts the inquiry: has the security concern regarding \textbf{MLaaS} for classification been effectively mitigated? 

The answer is negative. One severe security threat is model stealing, also known as model extraction  \cite{liang2022imitated,197128,DFME,DFMS}. In this process, an unauthorized party aims to replicate the decision of a target model by creating a substitute model. Attackers generate samples using a generative model, then acquire the target model's predictions for these samples and use them as supervisory information to iteratively refine the substitute model, until either the query budget is exhausted or the behavior of the substitute model closely matches that of the target model. Once the substitute model emulates the behavior of the target model, attackers can utilize it for malicious purposes, including adversarial attacks.
\begin{figure*}[!t]
    \centering
    \includegraphics[width=0.7\textwidth]{ 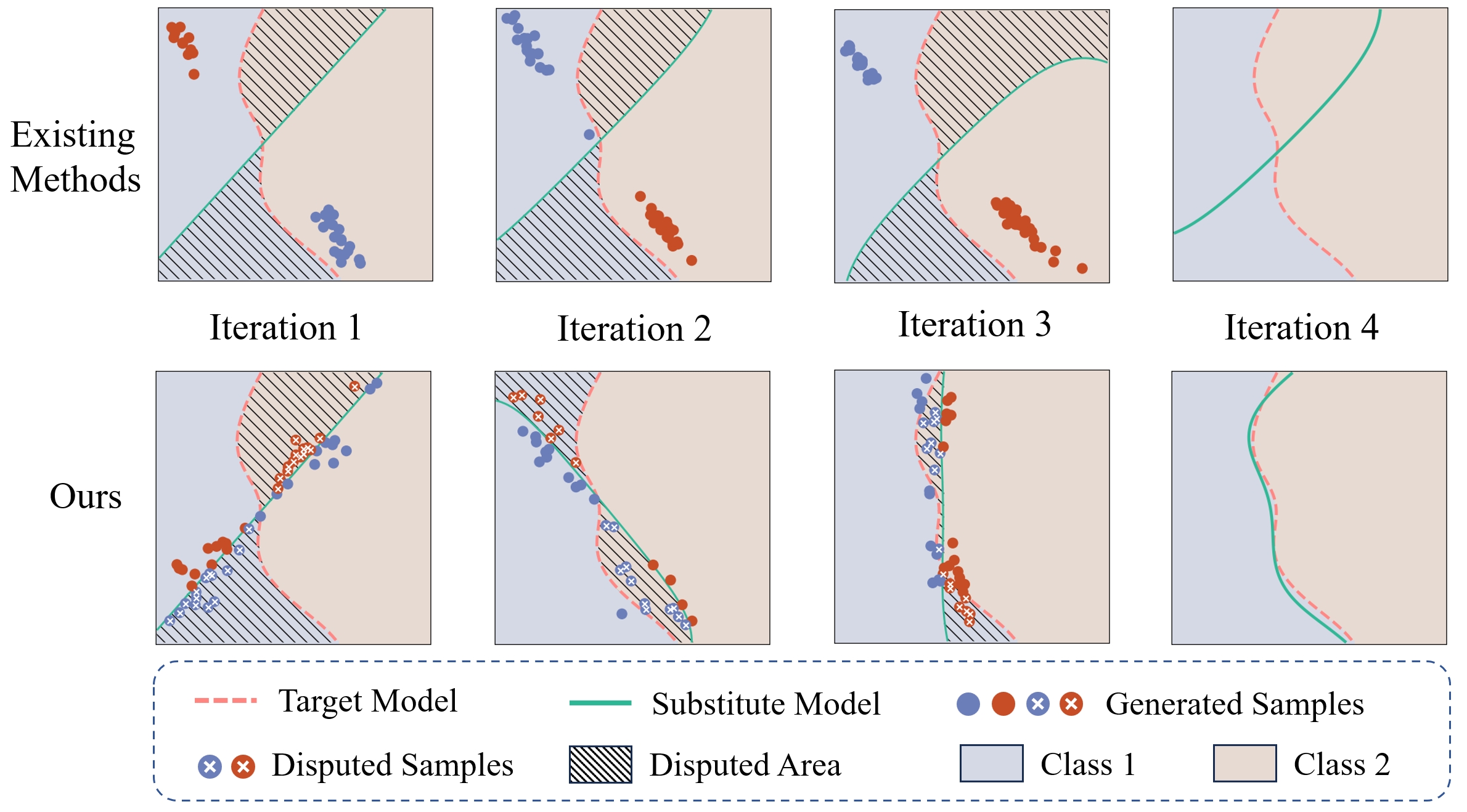}
     \caption{Compared to previous methods (top), our proposed approach (bottom) guides the generator to produce samples that are sufficiently close to the decision boundary of the substitute model. This results in more samples falling into the disputed area, thereby directing the substitute model to approach the target model in the correct direction.}
    \label{figure1}
\end{figure*}
The current methods within cooperative game frameworks for stealing models still have a significant issue \cite{DFTA22,ideal}. This can be understood by examining the entire process of model stealing. The attacker synthesizes features using pseudo labels via generator \cite{GAN}. As shown in the upper panel of Figure \ref{figure1}, the generator mainly produces synthetic features by minimizing the classification loss function of the substitute model. This strategy confers excessive confidence in the substitute model. The target and substitute models would have the same prediction of these synthetic features. Correspondingly, the synthetic features with target model predictions could not guide the update of the substitute model effectively. As a result, simulating the target model's decision-making requires a large number of synthetic features. Blindly increasing the number of queries to receive more feedback is unrealistic in real confrontation scenarios due to the risk of malicious behavior being detected \cite{prada}. Therefore, attackers should aim to generate the minimum number of synthetic features necessary, using these limited features to identify discrepancies between the target and substitute models. Ideally, these synthetic features should mainly occupy the disputed area, where the target and substitute models have different predictions of the same synthetic features.

To this end, we dissect the generation process of synthetic features and introduce a novel data-free model stealing method named Query Efficient Data Generation~(\textbf{QEDG}). Firstly, we implement a harmony loss to ensure the generated samples closely approach the classification boundary of multiple classes. Secondly, we apply a diversity loss by enlarging the gap between samples within the same class, thereby achieving a uniform distribution of generated samples along the decision boundary of the substitute model. Furthermore, we propose a query-free augmentation, which allows for the acquisition of more diverse supervised information without increasing the number of queries to the target model. The main contributions of this paper are summarized as follows:
\begin{itemize}
\item By dissecting the traditional synthetic process, we reconstruct the principal of the generator to ensure that the generated samples are close to and uniformly distributed along the decision boundary of the substitute model which not only causes some samples to fall into disputed areas, but also ensures that the generated samples cover as large a decision boundary surface as possible.
\item In order to make full use of the supervisory information obtained by interacting with the target model, a query-free sample augmentation is proposed to 
obtain multiple pieces of supervisory information through single query. Such a mechanism increases the diversity of samples.
\item Motivated by theoretical analysis, we introduce the consistency rate metric, which more accurately evaluates the similarity between the substitute and target models. We perform extensive experiments on four datasets using three metrics, where the results show that our method achieves better performance with a smaller number of queries.
\end{itemize}

\begin{figure*}[!t]
    \centering
    \includegraphics[width=0.7\textwidth]{ 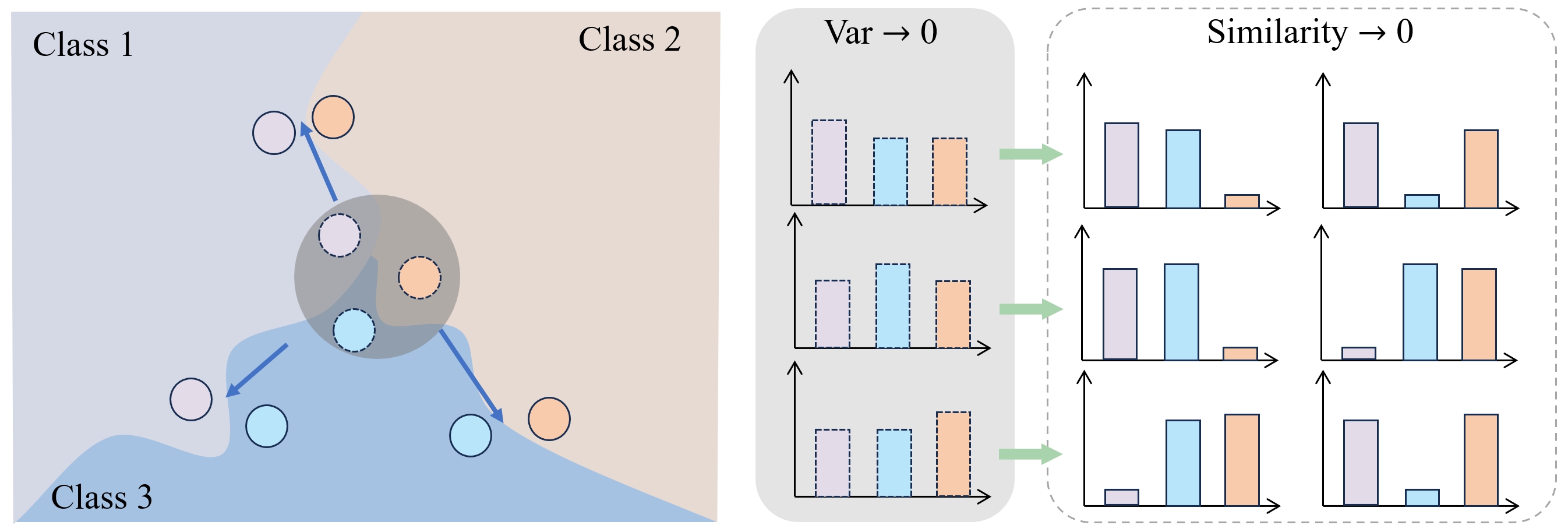}
    \caption{The confidence vectors of the points in the shaded area are displayed above. Our goal is to ensure that the generated sample points are as close to the decision boundary as possible while also minimizing intra-class similarity. This approach aims to distribute the generated samples as closely as possible along the decision boundary.}
    \label{method}
\end{figure*}
\section{Proposed Framework}
\subsection{Threat Modelling}
Here we will fully identify the potential threats of the adversary in data-free model stealing. 

\noindent\textbf{Goal.} To implement the adversarial attack with partial knowledge of the victim, the adversary aims to acquire a substitute model that could imitate the inference of the target model. Let $\boldsymbol{\theta}_T\in\mathbb{R}^{D_T}$ and $\boldsymbol{\theta}_S\in\mathbb{R}^{D_S}$ be the target and substitute model parameters, where $D_T\neq D_S$. Given $\boldsymbol{x}\in\mathbb{R}^D$ of a test sample $\boldsymbol{z}=\{\boldsymbol{x}, \boldsymbol{y}^*\}$, the adversary hopes 
\begin{equation}
   {T}(\boldsymbol{x}, \boldsymbol{\theta}_T) = {S}(\boldsymbol{x}, \boldsymbol{\theta}_S),
\end{equation}
where ${T}:\mathbb{R}^D\times\mathbb{R}^{D_T}\rightarrow\mathbb{R}$ is the decision function of the target model, and  ${S}:\mathbb{R}^D\times\mathbb{R}^{D_S}\rightarrow\mathbb{R}$ corresponds to the substitute model. It is noteworthy that the predictions of the target and substitute model could be inconsistent with the true label as $\boldsymbol{y}^*\neq{T}(\boldsymbol{x}, \boldsymbol{\theta}_T) = {S}(\boldsymbol{x}, \boldsymbol{\theta}_S)$. \\
\noindent\textbf{Knowledge.} Although $\boldsymbol{\theta}_T$ is agnostic to the model stealing attackers, he/she can get the feedback of the target model:
\begin{equation}
    \label{eq:true_label}
    \hat{\boldsymbol{y}}_g \overset{\underset{\mathrm{def}}{}}{=}{T}(\boldsymbol{x}_g, \boldsymbol{\theta}_T)
\end{equation}
with the synthetic sample $\boldsymbol{x}_g$ and construct $\boldsymbol{z}_g=\{\boldsymbol{x}_g,\hat{\boldsymbol{y}}_g\}$ to train the substitute model. Note that here we are considering a stricter scenario where the target model can only return the label, rather than a probability vector. The data-free attacker does not possess more knowledge than normal users in \textbf{MLaaS} scenario. 

\noindent\textbf{Ability.} The existing attackers \cite{DFTA22,ideal} generate $\boldsymbol{x}_g$ by minimizing the classification loss of the substitute model. This strategy confers excessive confidence in the substitute model. Such a behavior turns out to be a mirage. The target and substitute model could have the same prediction of $\boldsymbol{x}_g$ 
\begin{equation}
    \label{eq:opt_adv}
    {S}(\boldsymbol{x}_g, \boldsymbol{\theta}_S) = \hat{\boldsymbol{y}}_g.
\end{equation}
Yet the predictions of the substitute model would be still inconsistent with the target model on the true test samples
\begin{equation}
    \label{eq:inconsistent}
    {S}(\boldsymbol{x}, \boldsymbol{\theta}_S) \neq {T}(\boldsymbol{x}, \boldsymbol{\theta}_T).
\end{equation}

This issue requires a large number of synthetic samples $\{\boldsymbol{z}_g\}$ to intimate the decision boundary of the target model. The ``optimistic'' attackers need to continuously interact with the target model to query $\{\hat{\boldsymbol{y}}_g\}$ and construct $\{\boldsymbol{z}_g\}$. However, excessive interactions with the target model can lead to the attacker's samples $\boldsymbol{x}_g$ being identified as illegal inputs. In a real confrontation scenario, the adversary faces a paradox: he/she needs more feedback from the target model but the risk of his/her malicious behavior being detected increases. Different from the existing ``optimistic'' attackers, the proposed ``pessimistic'' adversary makes every effort to minimize the number of queries, \textit{i.e.} generate as few synthetic samples as possible. Consequently, the adversary pays attention to the synthetic $\boldsymbol{x}_g$ with different predictions from the target and substitute models 
\begin{equation}
    \label{eq:pess_adv}
    {S}(\boldsymbol{x}_g, \boldsymbol{\theta}_S) \neq \hat{\boldsymbol{y}}_g. 
\end{equation}

\subsection{A Parsimonious Generation Process}
The above threat modeling shows that minimizing the classification loss function of the substitute model is the main reason for the inefficiency of existing attack methods. Specifically, at the $t$ step, with the given pseudo label $\boldsymbol{y}\in\{0,1\}^K$ where $K$ is the number of classes, the existing methods adopt a classification loss function ($\mathcal{L}_{\text{clf}}$) like the cross-entropy to generate the synthetic feature
\begin{equation}
    \label{eq:gen_loss}
    \boldsymbol{x}_g^{(t)}\in\underset{\boldsymbol{x}}{\textbf{\textit{arg min}}}\ \mathcal{L}_{\text{clf}}\left({S}\left(\boldsymbol{x},\boldsymbol{\theta}^{(t)}_{S}\right), \boldsymbol{y}\right).
\end{equation}
Then the attacker queries the target model $T(\cdot)$ about $\boldsymbol{x}^{(t)}_g$. The parameter of the substitute model will be updated by $\boldsymbol{z}^{(t)}_g=\{\boldsymbol{x}_g^{(t)},\boldsymbol{y}^{(t)}_g\}$ as
\begin{equation}
    \label{eq:sub_loss}
    \boldsymbol{\theta}^{(t+1)}_{S}\in\underset{\boldsymbol{\theta}}{\textbf{\textit{arg min}}}\ \mathcal{L}_S\left({S}\left(\boldsymbol{x}_g^{(t)},\boldsymbol{\theta}\right), \hat{\boldsymbol{y}}^{(t)}_g\right),
\end{equation}
where $\mathcal{L}_S$ is the objective function of the substitute model. When \eqref{eq:opt_adv} holds with some occasional good synthetic samples, \eqref{eq:sub_loss} would update $\boldsymbol{\theta}_{S}$ very slowly. To reduce the number of interactions, the adversary needs to efficiently discover the differences between the target model and the substitute model. This inspires us to dissect the generation process of the synthetic features. The synthetic features generated by the cross-entropy function are separated by the substitute model $\boldsymbol{\theta}^{(t)}_{S}$. The corresponding predictions of the target and substitute models could be consistent (see the upper panel of Fig. 1). As a consequence, the ``pessimistic'' attacker tries to generate the disputed synthetic features with the given labels $\boldsymbol{y}$ and the current model parameter $\boldsymbol{\theta}^{(t)}_{S}$. The synthetic features $\boldsymbol{x}_g^{(t)}$ should lie as close as possible to the decision boundary of ${S}(\cdot,\boldsymbol{\theta}^{(t)}_{S})$. Such synthetic features have a high probability in the disputed area as \eqref{eq:pess_adv} if the substitute model can not imitate the target model.

We analyze the requirements of the desirable synthetic features through Figure \ref{method}. Generally speaking, the disputed example $\boldsymbol{x}^{(t)}_g$ for ${S}(\cdot)$ means that the prediction

\begin{equation}
    \begin{aligned}
    \label{eq:S}
        & S(\boldsymbol{x}^{(t)}_g) &\overset{\underset{\mathrm{def}}{}}{=}&\ \ {S}(\boldsymbol{x}^{(t)}_g, \boldsymbol{\theta}^{(t)}_{S})\\
        & &=&\ \ [s_1(\boldsymbol{x}^{(t)}_g),s_2(\boldsymbol{x}^{(t)}_g),\dots,s_K(\boldsymbol{x}^{(t)}_g)]
    \end{aligned} 
\end{equation}
lacks the discriminative capability. Treating $S(\boldsymbol{x}^{(t)}_g)$ as a group of random variables, we hope it holds a small variance. Let $\mathrm{Var}(S(\boldsymbol{x}^{(t)}_g))$ be the variance of $S(\boldsymbol{x}^{(t)}_g)$:
\begin{equation}
    \begin{aligned}
        \label{eq:var}
        & & &\ \ \mathrm{Var}\left(S\left(\boldsymbol{x}^{(t)}_g\right)\right)\\
        & &=&\ \ \frac{1}{K}\sum_{k=1}^{K}\left(s_k\left(\boldsymbol{x}^{(t)}_g\right)-\bar{S}\left(\boldsymbol{x}^{(t)}_g\right)\right)^2,
    \end{aligned}
\end{equation}
where $\bar{S}(\boldsymbol{x}^{(t)}_g)$ be the mean of $S(\boldsymbol{x}^{(t)}_g)$. Thus, we introduce \eqref{eq:var} as the loss function to emphasize the harmony of the synthetic features, as comes below:
\begin{equation}
    \label{eq:harm_loss}
    \mathcal{L}_{\text{harm}} = \frac{1}{N}\sum_{\boldsymbol{x}}\mathrm{Var}(S(\boldsymbol{x})),
\end{equation}
where $N$ is the total number of the synthetic features. However, simply minimizing \eqref{eq:harm_loss} can only generate the synthetic features that lie in the junction of different classes (the center of Figure \ref{method}). Furthermore, we need to generate the synthetic features along the decision boundaries between any two classes. To extend the range of synthetic features, we try to increase the diversity of the synthetic features belonging to the same class. Specifically, the other loss function is designed as follows:
\begin{equation}
    \label{eq:div_loss}
    \begin{aligned}
        & & &\ \ \mathcal{L}_{\text{div}}=-\frac{2}{N(N-1)} \sum_{i=1}^N\sum_{j=1,i\neq j}^N d(\boldsymbol{x}_i,\boldsymbol{x}_j),        
    \end{aligned}
\end{equation}
where $d:\mathbb{R}^D\times\mathbb{R}^D\rightarrow\mathbb{R}_{+}$ is the distance or dissimilarity function of the synthetic features. Here we choose to use cosine similarity. The final object function of the generator is a combination of \eqref{eq:gen_loss}, \eqref{eq:harm_loss}, and \eqref{eq:div_loss} as
\begin{equation}
    \label{eq:total loss}
    \mathcal{L}_G = \mathcal{L}_{\text{clf}} + \alpha\cdot\mathcal{L}_{\text{harm}} + \beta\cdot\mathcal{L}_{\text{div}},
\end{equation}
where $\alpha$ and $\beta$ are hyper-parameters. Our strategy for selecting hyper-parameters is to ensure that the magnitudes of the three different loss functions are as close to the same scale as possible.

\subsection{Query-free Augmentation and Substitute Model Training}
\begin{figure}[htbp]
    \centering
    \includegraphics[width=\columnwidth]{ 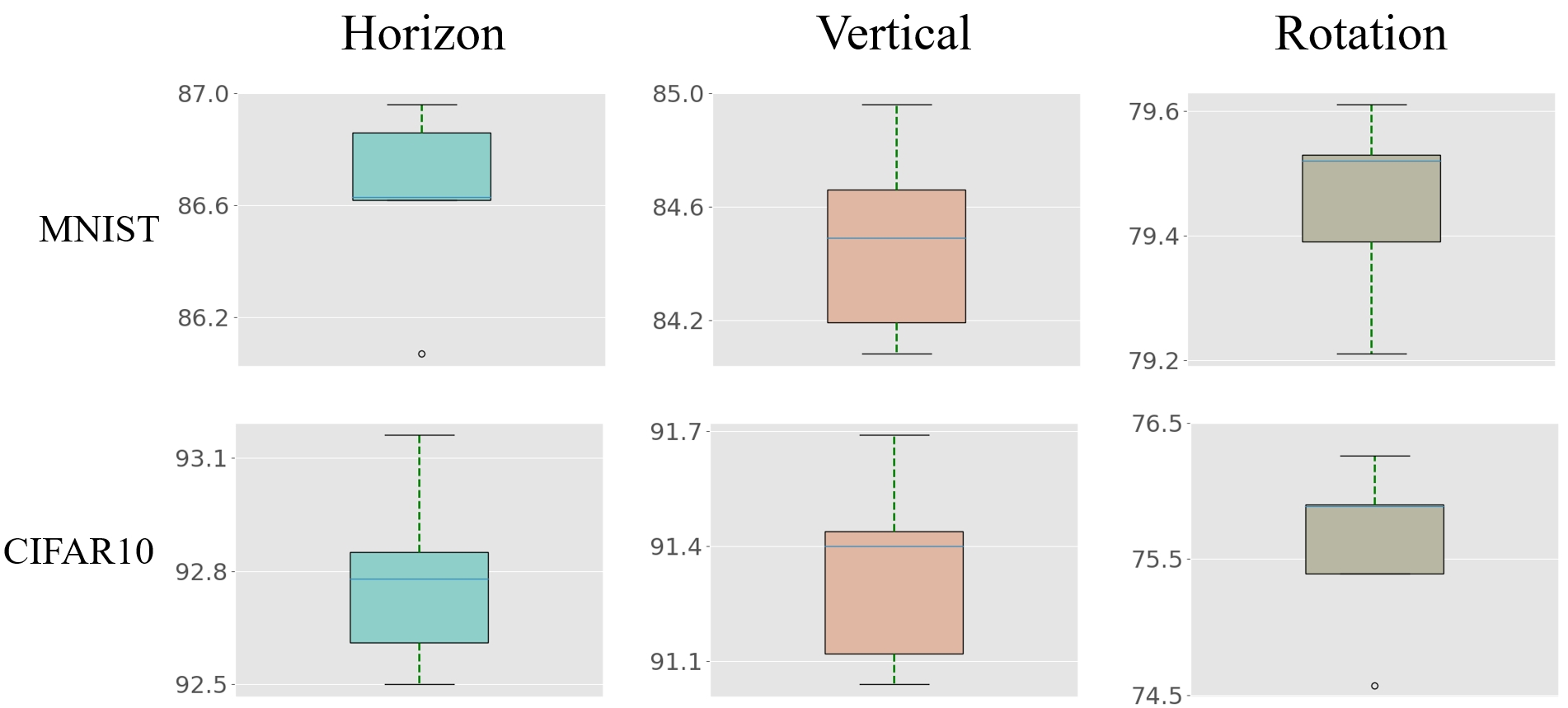}
     \caption{The consistency rate of predicted labels by the target model for non-disputed samples before and after data augmentation.}
    \label{aug}
\end{figure}
Once we obtain the synthetic data $\boldsymbol{x}_g$ using (\ref{eq:total loss}), we expect the outputs of the ${T}({\boldsymbol{x}_g})$ and ${S}({\boldsymbol{x}_g})$ to be as consistent as possible. 
However, using only one batch of features synthesized in the first stage of the current epoch, it will suffer from catastrophic forgetting  \cite{binici2022robust,do2022momentum}. The memory bank (\textbf{MB}) can effectively address this issue. Different from \cite{HEE} which only stores all previously synthesized features in the first stage, we also store the supervisory information from the target model. In the hard-label setting, the supervisory information refers to the label predicted by the target model for the features $\boldsymbol{x}_g$:
\begin{equation}
    \textbf{MB} = \textbf{MB} \cup \{(\boldsymbol{x}_g,\hat{\boldsymbol{y}}_g)\},\ \text{where}\ \hat{\boldsymbol{y}}_g = T(\boldsymbol{x}_g).
\end{equation}
We observe that HEE \cite{HEE} applies strong data augmentation to the features $\boldsymbol{x}_g$ stored in the memory bank. However, strong data augmentation can change the original labels. This raises the question: is it possible to achieve feature diversity without altering the original labels? We conducted experiments on both grayscale datasets like MNIST and colored images like CIFAR-10. For non-disputed samples, we applied three simple data augmentations, including horizontal flipping, vertical flipping, and rotation by a certain angle. In Figure \ref{aug}, we found that the labels of non-disputed samples remained high consistent rate before and after these data augmentations. The intuition behind this experimental phenomenon is that since non-disputed samples are correctly classified, they are further away from the decision boundary of the target model, making them more robust to simple data augmentations. Therefore, we retain the original labels of the non-disputed samples after data augmentation, thereby introducing more supervised information without additional query costs. It is worth noting that these augmented samples are not stored in the memory bank.

With the samples stored in the memory bank and the samples after random augmentation, we can update the parameters $\boldsymbol{\theta}_{S}$ of the substitute model. We assign different weights to the loss of disputed samples and non-disputed samples, respectively. The loss function of the substitute model $S(\cdot)$ at $t$ step is as follows:
\begin{equation}
    \label{eq:sub_loss2}
    \begin{aligned}
     & & &\ \ \mathcal{L}_{\mathcal{S}}^{(t)}(\boldsymbol{x}_g,\boldsymbol{y}_g;\boldsymbol{\theta})\\
     & &=&\left\{\begin{array}{rl}
        \gamma\cdot\mathcal{L}(\boldsymbol{x}_g,\boldsymbol{y}_g;\boldsymbol{\theta}), &\ \text{if}\ S\left(\boldsymbol{x}_g,\boldsymbol{\theta}_S^{(t-1)}\right)\neq\hat{\boldsymbol{y}}_g,\\
    \mathcal{L}(\boldsymbol{x}_g,\boldsymbol{y}_g;\boldsymbol{\theta}), &\ \text{otherwise},\\
    \end{array}\right. 
    \end{aligned}
\end{equation}
where 
\begin{equation}
    \mathcal{L}(\boldsymbol{x}_g,\boldsymbol{y}_g,\boldsymbol{\theta}) = \mathcal{L}_{\text{clf}}(S(\boldsymbol{x}_g,\boldsymbol{\theta}),\hat{\boldsymbol{y}}_g),
\end{equation}
and $\gamma$ is the hyper-parameter to control the importance of the disputed samples. The whole process of the proposed method is summarized as Algorithm \ref{algorithm-1}. 

\begin{algorithm}[htbp]
\caption{Query-efficient Data Generation}
\label{algorithm-1}
\KwIn{random noise $\boldsymbol{z}$, generator $G(\cdot)$, target model $T(\cdot)$, memory bank (\textbf{MB}), maximal number of queries $Q\geq1$, maximal iteration of generator $E\geq1$.}
\KwOut{$\boldsymbol{\theta}_{S}$.}
\textbf{Initialization:}\ $q=0,\ \textbf{MB}=\varnothing$;

\While{$q \leq Q$}{    
Given pseudo labels of a batch $\boldsymbol{Y}=\{\boldsymbol{y}_1,\boldsymbol{y}_2,\dots,\boldsymbol{y}_n\}$;

\For{$t=1,2,\dots, E$}{
$\boldsymbol{x}_{g}=G(\boldsymbol{z},\boldsymbol{\theta}_{G})$;

Update $\boldsymbol{\theta}_{G}$ using (12);
}
$\hat{\boldsymbol{y}}_g=T(\boldsymbol{x}_g)$ and Update $q$ ;

$\textbf{MB} = \textbf{MB} \cup \{(\boldsymbol{x}_{g},\hat{\boldsymbol{y}}_{g})\}$;

\For{$(\boldsymbol{x}_g,\hat{\boldsymbol{y}}_g) \in \textbf{MB}$}{
\eIf{$S(\boldsymbol{x}_g)==\hat{\boldsymbol{y}}_g$}{
$\boldsymbol{x}'_g=\text{Augmentation}(\boldsymbol{x}_g)$;

Update $\boldsymbol{\theta}_{S}$ using \eqref{eq:sub_loss2} with $(\boldsymbol{x}'_g,\hat{\boldsymbol{y}}_g)$;
}
{
Update $\boldsymbol{\theta}_{S}$ using \eqref{eq:sub_loss2} with $(\boldsymbol{x}_g,\hat{\boldsymbol{y}}_g)$;

}

}

}

\end{algorithm}

\subsection{Theoretical Analysis}
This part establishes the query complexity of the proposed active model stealing attack. Compared to the existing passive attackers, the following theorem states that the active attacker can easily steal any target model whose correct probability is greater than or equal to $1/2+c$ with some constant $c>0$ for all synthetic features. We show that the consistency rate between the target and substitute models increases by only a logarithmic multiplicative factor of the normal active learning query complexity. 

An active learning algorithm $(\boldsymbol{\mathcal{A}},\boldsymbol{\mathcal{T}})$ refers to an interaction between two players - the oracle $\boldsymbol{\mathcal{T}}$ and the learner $\boldsymbol{\mathcal{A}}$. $ \boldsymbol{\mathcal{A}}$ chooses $\boldsymbol{x}\in\boldsymbol{\mathcal{X}}$ and sends it to $\boldsymbol{\mathcal{T}}$, who responds with $\boldsymbol{y}\in\boldsymbol{\mathcal{Y}}$. $(\boldsymbol{x}, \boldsymbol{y})$ is then used by $\boldsymbol{\mathcal{A}}$ to obtain $\hat{f}$ from a hypothesis class $\boldsymbol{\mathcal{F}}$, which satisfies $\hat{f}(\boldsymbol{x})=\boldsymbol{y}$. We assume that the oracle $\boldsymbol{\mathcal{T}}$ adopts $f^*\in\boldsymbol{\mathcal{F}}$ to response $\boldsymbol{x}$ as $\boldsymbol{y}:=f^*(\boldsymbol{x})$. In this setting \cite{DBLP:conf/aaai/AlabdulmohsinGZ15,DBLP:conf/aaai/0003HK17,DBLP:conf/nips/Zhang0A20}, the different between $\hat{f}$ and $f^*$ is measured by 
\begin{equation}
\label{eq:15}
    \textbf{error}(\hat{f}) = \underset{\boldsymbol{x}\in\boldsymbol{\mathcal{X}}}{\boldsymbol{Pr}}\left\{\hat{f}(\boldsymbol{x})\neq f^*(\boldsymbol{x})\right\}.
\end{equation}
Then $q(\epsilon,\delta)$ denotes the query complexity of $\boldsymbol{\mathcal{A}}$ as $\boldsymbol{\mathcal{A}}$ chooses $\hat{f}\in\boldsymbol{\mathcal{F}}$ such that $\textbf{error}(\hat{f})\leq\epsilon$ with probability at least $1-\delta$.\\

\begin{theorem*}
Let $\boldsymbol{\mathcal{F}}$ be a hypothesis class and $(\boldsymbol{\mathcal{A}},\boldsymbol{\mathcal{T}})$ refer to an active learning algorithm as described above with the query complexity of $q(\epsilon,\delta)$. Suppose that an adversary $\boldsymbol{\mathcal{S}}$ disguises as $\boldsymbol{\mathcal{A}}$ but he/she can only receive imperfect feedback as 
\begin{equation}
    \label{impecf_feed}
    \phi(f^*,\boldsymbol{x})\overset{\underset{\mathrm{def}}{}}{=}\boldsymbol{Pr}\{\boldsymbol{y}_g\neq f^*(\boldsymbol{x})\}> 0,\ \forall\ \boldsymbol{x}\in\boldsymbol{\mathcal{X}}, 
\end{equation}
where $\boldsymbol{y}_g$ is the random variable that represents the feedback of $\boldsymbol{\mathcal{T}}$ to the $\boldsymbol{\mathcal{S}}$'s query $\boldsymbol{x}$. If $\textit{\textbf{max}}_{\boldsymbol{x}}\ \phi(f^*,\boldsymbol{x})<1/2$, $\boldsymbol{\mathcal{A}}$ could obtain $\hat{f}\in\boldsymbol{\mathcal{F}}$ such that
\begin{equation}
    \boldsymbol{Pr}\{\textbf{error}(\hat{f})\leq\epsilon\}\geq 1-2\delta,
\end{equation}
with query complexity 
\begin{equation}
    Q =\frac{8}{\left(1-2\cdot\underset{{\boldsymbol{x}}}{\textit{\textbf{max}}}\ \phi(f^*,\boldsymbol{x})\right)^2}\cdot q(\epsilon,\delta)\cdot\ln\frac{q(\epsilon,\delta)}{\delta}.
\end{equation}
\end{theorem*}
It is noteworthy that \eqref{impecf_feed} is a common situation for model stealing adversary. On the one hand, the oracle only exists is an ideal setting and the \textbf{MLaaS} API could provide some wrong predictions with the given queries. On the other hand, the potential defense mechanism would perturb the predictions of the target model to prevent model stealing. We provide the proof details in the supplementary materials. 
\begin{table}[ht]
\tabcolsep=0.08cm
\centering
\footnotesize
\caption{ASR(\%) comparisons between our method and baselines. We highlight the highest and second-highest values in \textcolor{orange}{orange} and \textcolor{cyan}{blue}, respectively. ``Untar.'' refers to non-targeted attacks, ``Tar.'' refers to targeted attacks.}
\label{table-asr-mnist&fmnist}
\begin{tabular}{cc ccc ccc ccc}
\toprule
\multicolumn{2}{c }{\multirow{2}{*}{Type}} & \multicolumn{2}{c }{\multirow{2}{*}{Method}} & \multirow{2}{*}{Query}&\multicolumn{3}{c }{MNIST} & \multicolumn{3}{c}{FMNIST}  \\
\multicolumn{2}{c }{} &  \multicolumn{2}{c }{} && FGSM & BIM & PGD & FGSM & BIM & PGD\\
\midrule
\multicolumn{2}{c }{\multirow{4}{*}{\rotatebox{90}{Untar.\quad\quad}}} 
&\multicolumn{2}{c }{HEE} 
&500K&\textcolor{cyan}{ 40.09} &\textcolor{cyan}{45.00} &\textcolor{cyan}{47.35 }&73.65 &67.46 &69.41\\
\multicolumn{2}{c }{\multirow{4}{*}{}} 
&\multicolumn{2}{c }{DaST} 
&250K& 12.40 &4.14 &5.68 &67.46 &29.26 &46.68\\
\multicolumn{2}{c }{\multirow{4}{*}{}}&\multicolumn{2}{c }{DFME} 
&250K& 27.12 &14.24 &15.85 &70.46 &47.16 &57.16\\
\multicolumn{2}{c }{\multirow{4}{*}{}}& \multicolumn{2}{c }{IDEAL} 
&20K&30.38 &14.35& 16.87 &75.76 &64.59& 70.88\\
\multicolumn{2}{c }{\multirow{4}{*}{}}& \multicolumn{2}{c }{DFTA} 
&20K&{34.67} &33.97& 39.19 &\textcolor{orange}{\textbf{83.40}} &\textcolor{cyan}{71.59}&\textcolor{cyan}{ 80.98}\\

\multicolumn{2}{c }{} & \multicolumn{2}{c }{\textbf{Ours}} &
\textbf{5K}&\textcolor{orange}{\textbf{41.37}}&\textcolor{orange}{\textbf{45.50}}& \textcolor{orange}{\textbf{50.21}} &\textcolor{cyan}{80.86}&\textcolor{orange}{\textbf{ 77.30}} &\textcolor{orange}{\textbf{81.33}}\\
\midrule
\multicolumn{2}{c }{\multirow{4}{*}{\rotatebox{90}{Tar.\quad\quad}}}
&\multicolumn{2}{c }{HEE} 
&500K&\textcolor{cyan}{ 15.50}&\textcolor{cyan}{15.34} &\textcolor{cyan}{16.29} &\textcolor{cyan}{14.79} &15.75 &15.86\\

\multicolumn{2}{c }{\multirow{4}{*}{}} 
&\multicolumn{2}{c }{DaST} 
&250K& 10.27 &9.97 &9.59 &9.97 &10.28 &9.48\\
\multicolumn{2}{c }{\multirow{4}{*}{}}&\multicolumn{2}{c }{DFME} 
&250K&12.09 & 11.23 & 11.76& 10.02 & 9.84 & 10.14 \\
\multicolumn{2}{c }{\multirow{4}{*}{}}& \multicolumn{2}{c }{IDEAL} 
&20K&12.80 &11.26& 11.85 &13.89 &12.57& 12.73\\
\multicolumn{2}{c }{\multirow{4}{*}{}}
&\multicolumn{2}{c }{DFTA} 
&20K&14.34 &13.15&14.15 &\textcolor{orange}{\textbf{16.11}}&\textcolor{cyan}{17.91} &\textcolor{cyan}{21.33}\\
\multicolumn{2}{c }{} & \multicolumn{2}{c }{\textbf{Ours}} &\textbf{5K}&\textcolor{orange}{\textbf{15.85}}& \textcolor{orange}{\textbf{16.62}} &\textcolor{orange}{\textbf{17.58}}&14.19&\textcolor{orange}{\textbf{21.42}}&\textcolor{orange}{\textbf{21.60}} \\
\bottomrule
\end{tabular}
\end{table}
\begin{table*}[!t]
\centering
\small
\setlength{\tabcolsep}{1.5mm} 
\caption{Accuracy(\%) and consistency(\%) comparisons between our proposed method and baselines over five datasets. We highlight the highest and second-highest values in \textcolor{orange}{orange} and \textcolor{cyan}{blue}, respectively. }
\label{table-con&acc}
\begin{tabular}{cc ccc ccc ccc ccc ccc}
\toprule
\multicolumn{2}{c }{\multirow{2}{*}{Method}} &\multicolumn{3}{c}{MNIST}&\multicolumn{3}{c}{FMNIST}& \multicolumn{3}{c}{SVHN}  & \multicolumn{3}{c}{CIFAR10} & \multicolumn{3}{c}{CIFAR100} \\
& & Query & Acc & Con & Query & Acc & Con & Query & Acc & Con& Query & Acc & Con & Query & Acc & Con\\
\midrule
\multicolumn{2}{c }{HEE} 
 &500k& 43.31&37.22&500k&27.07&19.26&1000k&39.75&31.36&1000k&34.82 &28.69&2M&15.87&16.73\\
\multicolumn{2}{c }{DaST} 
 &250k& 11.62&1.14&250k&16.43&5.48&250k&24.08&13.62&250k&11.92 &2.18&2M&17.26&19.38\\
\multicolumn{2}{c }{DFME} 
&250k &22.55 &12.38 &250k& 10.12 &1.38 &250k& 20.16 &9.09& 250k& 12.35& 2.94&2M&\textcolor{cyan}{20.17}&\textcolor{cyan}{22.14}\\
\multicolumn{2}{c }{IDEAL} 
&20k& 58.82& 54.26 &20k &22.48& 14.05& 50k&\textcolor{cyan}{60.71}&\textcolor{cyan}{54.93}& 50k &33.88 &27.47&400k&18.21&19.85\\
\multicolumn{2}{c }{DFTA} 
    &20k& \textcolor{cyan}{64.42}&\textcolor{cyan}{ 65.61} &20k &\textcolor{cyan}{46.95}& \textcolor{cyan}{41.48}& 50k&57.42 &51.68& 50k &\textcolor{cyan}{39.45} &\textcolor{cyan}{34.03}&400k&20.08&21.93\\
\multicolumn{2}{c }{\textbf{Ours}} 
&\textbf{5K}& \textcolor{orange}{\textbf{80.24 }}&\textcolor{orange}{\textbf{78.03}} &\textbf{5k}&\textcolor{orange}{\textbf{54.95}} &\textcolor{orange}{\textbf{50.86}} &\textbf{30K} &\textcolor{orange}{\textbf{66.58}} &\textcolor{orange}{\textbf{62.35}}&\textbf{30K} &\textcolor{orange}{\textbf{43.86}} &\textcolor{orange}{\textbf{39.45}}&\textbf{200k}&\textcolor{orange}{\textbf{22.32}}&\textcolor{orange}{\textbf{23.60}}\\
\bottomrule
\end{tabular}
\end{table*}
\begin{table*}[htbp]
\centering
\small
\caption{ASR(\%) comparisons between our method and baselines. We highlight the highest and second-highest values in \textcolor{orange}{orange} and \textcolor{cyan}{blue}, respectively. ``Untar.'' refers to non-targeted attacks. ``Tar.'' refers to targeted attacks.}
\label{table-asr-svhn&cifar10}
\resizebox{\textwidth}{!}{
\begin{tabular}{cc ccc ccc cccc cccc}
\toprule
\multicolumn{2}{c }{\multirow{2}{*}{Type}} & \multicolumn{2}{c }{\multirow{2}{*}{Method}} & \multirow{2}{*}{Query}&\multicolumn{3}{c }{SVHN} &  \multirow{2}{*}{Query}&\multicolumn{3}{c}{CIFAR10} &  \multirow{2}{*}{Query}&\multicolumn{3}{c}{CIFAR100} \\
\multicolumn{2}{c }{} &  \multicolumn{2}{c }{} && FGSM & BIM & PGD & &FGSM & BIM & PGD& &FGSM & BIM & PGD\\
\midrule
\multicolumn{2}{c }{\multirow{4}{*}{\rotatebox{90}{UnTar.\quad\quad}}}
&\multicolumn{2}{c }{HEE} 
&1000K& 72.48 &86.07 &86.17 &1000K&\textcolor{cyan}{72.12} &74.20 &72.36&2M&87.34&88.20&88.28\\
\multicolumn{2}{c }{\multirow{4}{*}{}} 
&\multicolumn{2}{c }{DaST} 
&250K& 58.69 &46.80 &49.22  &250K&58.61 &43.48 &47.24&2M&91.49&91.59&92.40\\
\multicolumn{2}{c }{\multirow{4}{*}{}}&\multicolumn{2}{c }{DFME} 
&250K& 62.72 &67.33 &67.20  &250K&59.76 &48.99 &50.29&2M&92.07&\textcolor{cyan}{91.75}&\textcolor{cyan}{92.99}\\
\multicolumn{2}{c }{\multirow{4}{*}{}}& \multicolumn{2}{c }{IDEAL} 
&50K& \textcolor{orange}{\textbf{78.93}}&85.46& 84.51  &50K&68.24 &70.58& 70.86&400K&91.66&89.24&91.15\\
\multicolumn{2}{c }{\multirow{4}{*}{}}& \multicolumn{2}{c }{DFTA} 
&50K&72.14 &\textcolor{cyan}{86.95} &\textcolor{cyan}{87.30}  &50K&71.82 &\textcolor{cyan}{75.97}& \textcolor{cyan}{78.16}&400K&\textcolor{orange}{\textbf{92.20}}&90.17&91.16\\
\multicolumn{2}{c }{} & \multicolumn{2}{c }{\textbf{Ours}} &
\textbf{30K}&\textcolor{cyan}{73.90}&\textcolor{orange}{\textbf{89.31}}& \textcolor{orange}{\textbf{87.83}}  &\textbf{30K}&\textcolor{orange}{\textbf{72.95}}& \textcolor{orange}{\textbf{76.51}} &\textcolor{orange}{\textbf{78.64}}&\textbf{200K}&\textcolor{cyan}{91.94}&\textcolor{orange}{\textbf{91.84}}&\textcolor{orange}{\textbf{93.07}}\\
\midrule
\multicolumn{2}{c }{\multirow{4}{*}{\rotatebox{90}{Tar.\quad\quad}}}
&\multicolumn{2}{c }{HEE} 
&1000K& \textcolor{orange}{\textbf{22.22}} &55.64 &55.78 &1000K &\textcolor{orange}{\textbf{21.87}} &\textcolor{cyan}{31.23} &29.73&2M&\textcolor{orange}{\textbf{2.04}}&\textcolor{cyan}{12.03}&\textcolor{cyan}{11.45}\\

\multicolumn{2}{c }{\multirow{4}{*}{}} 
&\multicolumn{2}{c }{DaST} 
&250K&11.40&22.24 &21.85 &1000K &9.62 &10.22 &10.56&2M&1.63&10.88&10.47\\
\multicolumn{2}{c }{\multirow{4}{*}{}}&\multicolumn{2}{c }{DFME} 
&250K&10.42& 12.56 & 12.55  &1000K& 10.65 & 10.27 & 10.42 &2M&1.61&11.26&10.34\\
\multicolumn{2}{c }{\multirow{4}{*}{}}&\multicolumn{2}{c }{IDEAL} 
&50K&\textcolor{cyan}{20.35} &\textcolor{cyan}{62.26} &\textcolor{cyan}{58.20}&50K&15.11&30.49 &\textcolor{cyan}{30.83}&400K&\textcolor{cyan}{1.83}&9.04&8.81\\
\multicolumn{2}{c }{\multirow{4}{*}{}}&\multicolumn{2}{c }{DFTA} 
&50K&17.60 &54.96 &51.76&50K &14.95&29.96 &27.93&400K&1.68&9.96&9.63\\
\multicolumn{2}{c }{} & \multicolumn{2}{c }{\textbf{Ours}} &\textbf{30K}&18.87& \textcolor{orange}{\textbf{64.28}} &\textcolor{orange}{\textbf{61.19}} &\textbf{30K}&\textcolor{cyan}{15.46}&\textcolor{orange}{\textbf{31.88}}&\textcolor{orange}{\textbf{31.33}} &200K&1.68&\textcolor{orange}{\textbf{12.66}}&\textcolor{orange}{\textbf{12.08}}\\
\bottomrule
\end{tabular}
}
\end{table*}
\section{Experiments}
\subsection{Experiment Setup}
\label{Experiment Setup}
\textbf{Datasets and Model Architectures} \quad We conducted experiments on five well-known datasets: MNIST, FMNIST, SVHN, CIFAR10, CIFAR100. We unified the target models of different methods, as well as the substitute models of different methods, noting that the target models and substitute models are different. We separately tested different model structures, including ResNet   \cite{resnet}, VGG  \cite{vgg}, and our own designed CNN model. Furthermore, we adopted the architecture described in  \cite{DFTA22} and employed the same generator as used in StyleGAN  \cite{Style}.\\
\textbf{Baselines} \quad We compare our approach with the following state-of-the-art (SOTA) methods. DaST  \cite{DaST} and DFME  \cite{DFME} and HEE \cite{HEE} are three data-free model stealing methods that do not take into account the number of queries. DFTA  \cite{DFTA22} and IDEAL  \cite{ideal} are query efficient and achieve improvements over the above methods. \\
\textbf{Evaluation Metrics}\quad We employ three metrics for evaluation. Accuracy measures the precision with which the substitute model performs predictions on the actual test set. Motivated by theoretical analysis and (\ref{eq:15}), we also consider consistency metric. To further exclude the possibility that agreement may occur by chance between target model and substitute model. we adopt Cohen's kappa \cite{ck}. It measures the agreement between target model and substitute model. This metric is not affected by the accuracy of the target model. The calculation formula for Cohen's kappa is as follows:
\begin{equation}
    \kappa = \frac{p_o-p_e}{1-p_e}
\end{equation}
where $p_e$ represents the relative consistency between the two models and is calculated by summing of the products of the marginal probabilities for each class. $p_o$ represents the accidental consistency between the two models and is calculated by dividing the sum of the diagonal elements of the confusion matrix by the total number of samples. To further test the consistency between the substitute model and the target model, we also test the attack success rate. We employ three common attack methods to generate adversarial examples: FGSM   \cite{FGSM}, BIM  \cite{BIM}, and PGD  \cite{PGD}. \\
\textbf{Implementation details}
\quad We train the substitute model using an SGD optimizer with a batch size of 256, an initial learning rate of 0.01, a momentum of 0.9, and no weight decay. For the generator, we adopt a batch size of 256, utilizing an adam optimizer  \cite{ adam} with a fixed learning rate of 0.001. For both MNIST and FMNIST, the perturbation bound is respectively set to 0.3, 0.2, and 0.1 with a step size $\alpha$ of 0.01. For SVHN, CIFAR10 and CIFAR100, the perturbation bound is set to 32/255, 20/255, 8/255 with a step size $\alpha$ of 0.01. In untargeted attacks, we only attack images correctly classified by the target model. In targeted attacks, we only attack images correctly classified by the target model with labels different from their original labels. This is as the same as DFTA \cite{DFTA22}. 
\subsection{Experiment Results}
\begin{figure}[!t]
    \centering
    \includegraphics[width=0.75\columnwidth]{ 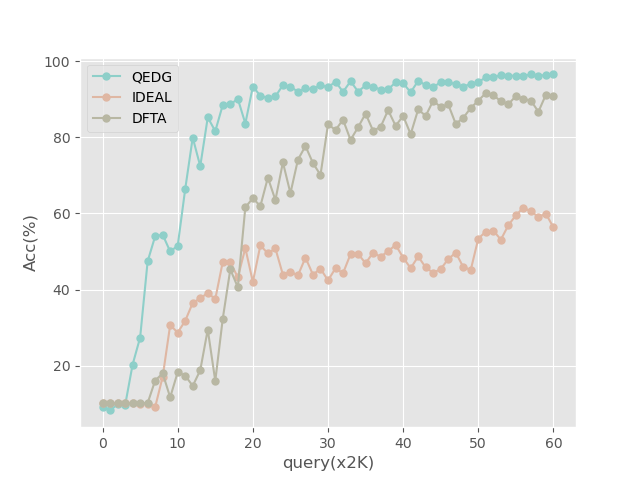}
     \caption{The accuracy of our method and other methods varies with the number of queries on Microsoft Azure.}
    \label{Azure}
\end{figure}

\noindent\textbf{Attacks on the Microsoft Azure Online Model\footnote{https://www.customvision.ai}}\quad We conducted experiments in real-world scenarios using Microsoft's Azure API service. This service restricts users to uploading their own datasets, and we utilized the MNIST dataset for tests, consistent with  \cite{ideal}. The architecture and parameters of the model provided by the service are not disclosed. For the substitute model, we employed ResNet18. From the Figure \ref{Azure}, we can observe that our method achieves an accuracy comparable to DFTA's accuracy at 100k queries when using only 20k queries. Moreover, with a sufficient number of queries, our method is capable of achieving even higher accuracy.

\noindent\textbf{Experiments on MNIST and FMNIST}\quad We used ResNet18 as the substitute model and ResNet34 as the target model. We tested three evaluation metrics: consistency rate, accuracy, and attack success rate. The perturbation bound was set to 0.3. The attack success rates for other perturbation bounds can be found in the appendix. We also tested results when both the target and substitute models were CNN structures. These results are also in the appendix. Table \ref{table-con&acc} shows the accuracy and consistency rate.
Our method outperforms others by a significant margin in both accuracy and consistency rate, even with fewer queries. This is because our method generates more samples in the disputed region, enabling each query to effectively guide the substitute model's update. In contrast, other methods generate overly confident samples for the substitute model. Few samples fall into the disputed area, resulting in significant query waste.
We test the success rates of targeted and untargeted attacks using three attack methods. Table \ref{table-asr-mnist&fmnist} shows that regardless of the attack method, our method achieves the highest attack success rates with the fewest queries. This indicates that the decision boundary learned by our substitute model is most similar to the target model's.

\noindent\textbf{Experiments on SVHN, CIFAR10 and CIFAR-100}\quad 
%
Due to the simplicity of grayscale images, we extended our experiments to more complex datasets like SVHN and CIFAR10, CIFAR100 which consist of color images. We used the same model structure and evaluation metric. The perturbation bound was set to 32/255. The attack success rates for other perturbation bounds can be found in the appendix. We also tested results with the target model set to VGG19. These results are also in the appendix. We used more queries on both datasets because the decision boundaries of the target models are more complex. Table \ref{table-con&acc} shows that our method surpasses others in consistency rate and accuracy, requiring only 30k queries compared to 50k or even 250k needed by other methods. In addition, we also test the transfer attack success rate under different attack methods. From the table \ref{table-asr-svhn&cifar10}, it can be seen that our method achieves a higher transfer attack success rate in most cases with fewer query number.
We also plotted line charts showing the relationship between accuracy, consistency, and the number of queries in the appendix. Figure shows that the accuracy and consistency of other methods fluctuate, while ours steadily increase.\\ 
\begin{figure}[htbp]
    \centering
    \includegraphics[width=1\columnwidth]{ 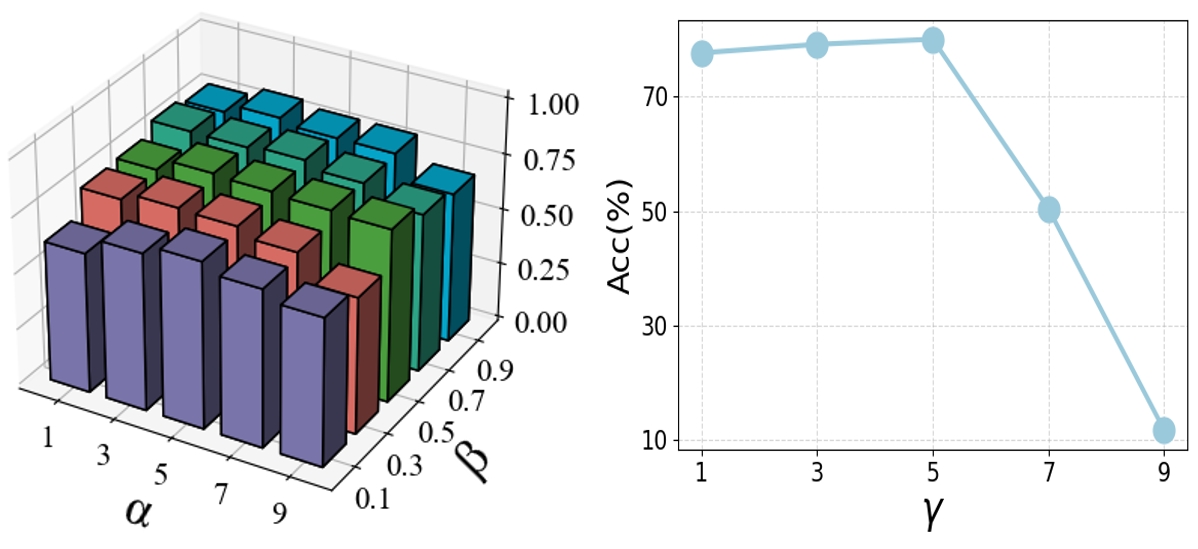}
     \caption{Sensitivity analysis about hyperparameters $\alpha$ and $\beta$ and $\gamma$.}
    \label{hyper}
\end{figure}
\noindent\textbf{Sensitivity Analysis}\quad We conducted a hyperparameter analysis on the MNIST dataset and found that optimal performance can only be achieved when the two loss functions are on the same order of magnitude. We set $\alpha=5$ and $\beta=0.7$. Regarding the selection of $\gamma$, the experiments revealed that assigning a higher weight to contentious samples can improve the performance of our method. However, if the weight is too high, the performance will quickly decline. Here we set $\gamma=5$.

\subsection{Ablation Study}
\begin{table}[H]
\centering
\caption{Ablation study on the proposed components. The highest score on each column is shown in \textbf{bold}.}
\label{table-ablation}
\resizebox{0.3\textwidth}{!}{
\begin{tabular}{cccccc}
\toprule
\multirow{2}{*}{QA} & \multirow{2}{*}{$\mathcal{L}_{\text{harm}}$} &\multirow{2}{*}{$\mathcal{L}_{\text{div}}$} & \multicolumn{3}{ c }{CIFAR10} \\
\multirow{2}{*}{} & \multirow{2}{*}{} &\multirow{2}{*}{} & ACC & Con & ASR\\ 
\midrule
\Checkmark &  & &31.9&25.8&69.5 \\
 & \Checkmark &   &23.1 &14.8& 67.9\\
  &   & \Checkmark &40.8&36.0&70.3\\
\Checkmark & \Checkmark &   &31.2 &25.7&69.3\\
\Checkmark &   &\Checkmark  &41.9&37.3&71.6 \\
  & \Checkmark & \Checkmark &38.8&33.4&70.5\\
\Checkmark & \Checkmark & \Checkmark &\textbf{43.8} & \textbf{39.4} & \textbf{72.9}\\
\bottomrule
\end{tabular}}
\end{table}

In this section, we conducted comprehensive ablation experiments on CIFAR10 to assess the significance of each component of our method, with the results presented in Table~\ref{table-ablation}. Our approach primarily consists of three improvement components: query-free augmentation~(QA), $\mathcal{L}_{\text{harm}}$, and $\mathcal{L}_{\text{div}}$. We tested the experimental effects of adding or removing each module separately, resulting in a total of six sets of experiments. Under otherwise identical conditions, using $\mathcal{L}_{\text{div}}$ yielded higher performance compared to using QA and $\mathcal{L}_{\text{harm}}$, and we achieved the best performance when all three components are used simultaneously. Because $\mathcal{L}_{\text{harm}}$ prefers the centers of decision surfaces with multiple classes, if only $\mathcal{L}_{\text{harm}}$ appears without $\mathcal{L}_{\text{div}}$, it leads to negative optimization of the loss function. This is evidenced by the experiments, where the results with only $\mathcal{L}_{\text{harm}}$ and without $\mathcal{L}_{\text{div}}$ are the worst. The role of QA is to increase the diversity of samples without increasing the number of queries. Experimental results also indicate that having QA can significantly improve the model's performance.

\section{Conclusion and Limitations}
To reduce the number of queries required in model stealing, we dissect the existing synthetic process and reconstruct the principal of the generator to ensure that the synthetic samples are close to and uniformly distributed along the decision boundary of the substitute model. The generated samples should obtain different predictions of the target and substitute models and cover as large a decision boundary surface as possible. Furthermore, a query-free sample augmentation is proposed to make full use of the supervisory information obtained by interacting with the target model. Then we establish the query complexity of the proposed active adversary with imperfect feedback of the target model. We further introduce the consistency rate metric, which more accurately evaluates the similarity between the substitute and target models. Our method outperformed the current state-of-the-art competitors on the real \textbf{MLaaS} scenario and five datasets. However, both our method and existing approaches generate samples that lack any semantic information.
\bibliography{aaai25}
\appendix
\onecolumn

\section{Related Work}
Deep learning has been applied to various domains, but it faces a wide range of security challenges including poisoning attack \cite{ma2021poisoning}, sequential manipulation\cite{ma2024sequential,ma2022tale}, backdoor attack\cite{liu2024does,liang2024badclip}, adversarial attack and model stealing, etc.
\subsection{Black-box attack}
Black-box attacks refer to situations where attackers construct adversarial samples without accessing the model's internal structure and parameters. Black-box attacks are mainly of two types: query-based and transfer-based. Query-based attacks can be divided into score-based attacks and decision-based attacks. Score-based attacks \cite{liang2022large,ZOO,simple,square,liang2021parallel} update adversarial samples by observing changes in loss indicated by DNN’s output scores, such as logits or probabilities. However, score-based attacks can be defended by injecting parametric noise\cite{pni} or slightly modifying the output logits to mislead the attack\cite{NEURIPS2022_5fa29a2f}. In real-life scenarios, MLaaS usually returns only the top-1 labels instead of logits or probabilities. Decision-based attacks rely only on DNN’s decisions, such as top-1 labels, to generate adversarial examples\cite{bda}. Since decision-based attacks cannot perform greedy updates, they start with a different sample and aim to keep the DNN’s prediction wrong, requiring thousands of queries to achieve a non-trival success rate. The other type is transfer-based attacks\cite{jia2022adversarial,admix,TVT,tid,gu2023survey,gao2025boosting,jia2023transegpgd}, which rely on the transferability of adversarial samples. They involve training a substitute model on a training dataset and using its white-box characteristics to construct adversarial samples to attack the target model. However, due to data privacy and commercial value concerns, MLaaS providers may not disclose their training datasets.

\subsection{Data-free Model Stealing}
In real-world scenarios, due to data privacy and commercial value, MLaaS providers do not disclose their training datasets, and the models' structures and parameters are opaque. MLaaS providers also only return the top-1 labels for user inputs. This scenario is both more challenging and more practical. Therefore, many methods now aim to steal target models without any prior knowledge about the target model. These methods can be broadly categorized into two types. The first type \cite{DFMS,Active-Thief,EDST} tries to steal the target model's functionality using proxy data. However, obtaining proxy data isn't always easy. If there are significant differences between the proxy data and the target model's training data, it can affect the substitute model's training. The second type does not use any real data \cite{chandrasekaran2020exploring}. It generates a synthetic dataset using generators to create fake data from noise. DaST \cite{DaST} was the first to steal the functionality of a black-box model without real data. However, the generator size increases significantly as the class number of the victim dataset grows. DDG \cite{Delving} then modified the generator architecture to compress its size and explore more effective data. These methods model the generator and substitute model as a zero-sum game, requiring many queries. DFME \cite{DFME} and MAZE\cite{Maze} estimate the black-box using zero-order gradient estimation \cite{ZOO} to update the generator. Gradient estimation also requires many queries. More queries make an attacker easier to detect. DFTA\cite{DFTA22} and IDEAL\cite{ideal} transforms the zero-sum game into a cooperative game, reducing the number of queries.
In this paper, we explore guiding the generator to produce samples inconsistent with both the substitute and target models' predictions within a cooperative game framework, using fewer queries. This allows for more efficient updates of the substitute model.
       
\section{Theoretical Analysis}
\begin{theorem*}
Let $\boldsymbol{\mathcal{F}}$ be a hypothesis class and $(\boldsymbol{\mathcal{A}},\boldsymbol{\mathcal{T}})$ refer to an active learning algorithm as described above with the query complexity of $q(\epsilon,\delta)$. Suppose that an adversary $\boldsymbol{\mathcal{S}}$ disguises as $\boldsymbol{\mathcal{A}}$ but he/she can only receive imperfect feedback as 
\begin{equation}
    \label{impecf_feed}
    \phi(f^*,\boldsymbol{x})\overset{\underset{\mathrm{def}}{}}{=}\boldsymbol{Pr}\{\boldsymbol{y}_g\neq f^*(\boldsymbol{x})\}> 0,\ \forall\ \boldsymbol{x}\in\boldsymbol{\mathcal{X}}, 
\end{equation}
where $\boldsymbol{y}_g$ is the random variable that represents the feedback of $\boldsymbol{\mathcal{T}}$ to the $\boldsymbol{\mathcal{S}}$'s query $\boldsymbol{x}$. If $\textit{\textbf{max}}_{\boldsymbol{x}}\ \phi(f^*,\boldsymbol{x})<1/2$, $\boldsymbol{\mathcal{A}}$ could obtain $\hat{f}\in\boldsymbol{\mathcal{F}}$ such that
\begin{equation}
    \boldsymbol{Pr}\{\textbf{error}(\hat{f})\leq\epsilon\}\geq 1-2\delta,
\end{equation}
with query complexity 
\begin{equation}
    Q =\frac{8}{\left(1-2\cdot\underset{{\boldsymbol{x}}}{\textit{\textbf{max}}}\ \phi(f^*,\boldsymbol{x})\right)^2}\cdot q(\epsilon,\delta)\cdot\ln\frac{q(\epsilon,\delta)}{\delta}.
\end{equation}
\end{theorem*}

\begin{proof}
    Here we focus on the binary case as $\boldsymbol{\mathcal{Y}}=\{0,1\}$ and the results are easily extended to the multi-class case. First, we clarify the behavior of the adversary $\boldsymbol{\mathcal{S}}$. Follow Algorithm \ref{algorithm-1}, $\boldsymbol{\mathcal{S}}$ will
    \begin{itemize}
        \item generate the synthetic feature $\boldsymbol{x}_g$ with the generator $G(\boldsymbol{z},\boldsymbol{\theta}_G,\boldsymbol{y}_k)$ using loss function like \eqref{eq:total loss}, where $\boldsymbol{z}$ is the random noise, $\boldsymbol{y}_k$ is the given pseudo label;
        \item query $\boldsymbol{x}_g$ to $\boldsymbol{\mathcal{T}}$ and obtain $\hat{\boldsymbol{y}}_g=f^*(\boldsymbol{x}_g)$, which could be the imperfect feedback;
        \item use $\{(\boldsymbol{x}_g,\hat{\boldsymbol{y}}_g)\}$ to learn the substitute model $\hat{f}$.
    \end{itemize}
    By the Boole's inequality, \textit{a.k.a} union bound, it holds that 
    \begin{equation}
        \label{eq:24}
        \boldsymbol{Pr}\left\{\textbf{error}(\hat{f})\leq\epsilon\right\}\geq1-\delta-\left(1-\boldsymbol{Pr}\left\{\bigcap_{t=1}^{|\{(\boldsymbol{x}_g,\hat{\boldsymbol{y}}_g)\}|}\left\{\hat{\boldsymbol{y}}_g=f^*(\boldsymbol{x}_g)\right\}\right\}\right). 
    \end{equation}
    Without loss of generality, let the cardinality of $\{(\boldsymbol{x}_g,\hat{\boldsymbol{y}}_g)\}$ be $R$ times of the active learning sample complexity $q(\epsilon, \delta)$ for $(\boldsymbol{\mathcal{A}},\boldsymbol{\mathcal{T}})$ as
    \begin{equation}
        \left|\left\{(\boldsymbol{x}_g,\hat{\boldsymbol{y}}_g)\right\}\right| = R\cdot q(\epsilon, \delta)
    \end{equation}
    where $R$ is the maximal interaction times of a synthetic feature $\boldsymbol{x}_g$ and the duplicated interactions arise from data augmentation and the imperfect feedback. Furthermore, we introduce a random variable $\tau^r_g$ as 
    \begin{equation}
        \tau^r_g = \left\{\begin{array}{cl}
            1, &\text{if}\ f^*(\boldsymbol{x}_g) = \boldsymbol{y}_g\\
            0, &\text{otherwise,} 
        \end{array}
        \right.
    \end{equation}
    and
    \begin{equation}
        \tau_g = \sum_{r=1}^R \tau^r_g. 
    \end{equation}
    Then we know that 
    \begin{equation}
        \boldsymbol{Pr}\left\{\bigcap_{g=1}^{|\{(\boldsymbol{x}_g,\hat{\boldsymbol{y}}_g)\}|}\left\{\hat{\boldsymbol{y}}_g=f^*(\boldsymbol{x}_g)\right\}\right\}\geq\boldsymbol{Pr}\left\{\bigcap_{g=1}^{q(\epsilon, \delta)}\left\{\tau_g>\frac{R}{2}\right\}\right\}\geq 1 - \sum_{g=1}^{q(\epsilon, \delta)}\boldsymbol{Pr}\left\{\tau_g\leq\frac{R}{2}\right\}
    \end{equation}
    where the last step also follows the Boole's inequality. With the Chernoff inequality, we have 
    \begin{equation}
        \boldsymbol{Pr}\left\{\tau_g\leq\frac{R}{2}\right\}\leq \textbf{\textit{exp}}\left(-\frac{R}{2}\left(\underset{{\boldsymbol{x}_g}}{\textit{\textbf{max}}}\ \phi(f^*,\boldsymbol{x}_g)-1/2\right)^2\right).
    \end{equation}
    As a consequence, \eqref{eq:24} has a lower bound as
    \begin{equation}
        \boldsymbol{Pr}\left\{\textbf{error}(\hat{f})\leq\epsilon\right\}\geq1-\delta-q(\epsilon,\delta)\cdot\textbf{\textit{exp}}\left(-\frac{R}{2}\left(\underset{{\boldsymbol{x}_g}}{\textit{\textbf{max}}}\ \phi(f^*,\boldsymbol{x}_g)-1/2\right)^2\right).
    \end{equation}
    By setting
    \begin{equation}
        R = \frac{8}{\left(1-2\cdot\underset{{\boldsymbol{x}}}{\textit{\textbf{max}}}\ \phi(f^*,\boldsymbol{x})\right)^2}\cdot \ln\frac{q(\epsilon,\delta)}{\delta},
    \end{equation}
    we know that the adversary $\boldsymbol{\mathcal{S}}$ could implements the model stealing and archive 
    \begin{equation}
        \boldsymbol{Pr}\{\textbf{error}(\hat{f})\leq\epsilon\}\geq 1-2\delta
    \end{equation}
    with the query complexity 
    \begin{equation}
        Q =\frac{8}{\left(1-2\cdot\underset{{\boldsymbol{x}}}{\textit{\textbf{max}}}\ \phi(f^*,\boldsymbol{x})\right)^2}\cdot q(\epsilon,\delta)\cdot\ln\frac{q(\epsilon,\delta)}{\delta},
    \end{equation}
    which is our claim.
\end{proof}

\end{document}